\documentclass[a4paper,10pt]{article}
\usepackage{graphicx} 

\usepackage[utf8]{inputenc} 
\usepackage{authblk}        
\usepackage{amsmath}
\usepackage{tabularx}
\usepackage{amssymb}
\usepackage{setspace}
\usepackage{multirow}
\usepackage{graphicx}
\usepackage[table,usenames,dvipsnames]{xcolor}
\usepackage{booktabs} 

\usepackage[cal=boondoxo]{mathalfa} 
\usepackage{amssymb}
\usepackage{setspace}
\usepackage{multirow}

\newcommand{\tb}[1]{\textcolor{blue}{#1}}
\renewcommand{\tb}[1]{#1}

\newcommand{\avg}[1]{\langle#1\rangle}

\title{Interpreting Node Embedding Distances\\Through $n$-order Proximity Neighbourhoods \footnote{This is a preprint of the following chapter: Shakespeare et al., Interpreting Node Embedding Distances Through $n$-order Proximity Neighbourhoods, published in Complex Networks XV, edited by Federico Botta, Mariana Macedo, Hugo Barbosa, Ronaldo Menezes, 2024, Springer Cham reproduced with permission of Springer Cham. The final authenticated version is available online at: https://doi.org/10.1007/978-3-031-57515-0}}

\author[1,2]{Dougal Shakespeare}
\author[1,2]{Camille Roth}

\affil[1]{CNRS (Centre National de la Recherche Scientifique), France}
\affil[2]{CAMS (Centre d'Analyse et de Mathématique Sociale), EHESS / CNRS, Paris}

\begin{document}
\date{}

\maketitle

\begin{abstract}
In the field of node representation learning the task of interpreting latent dimensions has become a prominent, well-studied research topic. The contribution of this work focuses on appraising the interpretability of another rarely-exploited feature of node embeddings increasingly utilised in recommendation and consumption diversity studies: inter-node embedded distances. Introducing a new method to measure how understandable the distances between nodes are, our work assesses how well the proximity weights derived from a network before embedding relate to the node closeness measurements after embedding. Testing several classical node embedding models, our findings reach a conclusion familiar to practitioners albeit rarely cited in literature - the matrix factorisation model SVD is the most interpretable through 1, 2 and even higher-order proximities.
\end{abstract}

\section{Introduction}\label{sec:headings}
In the field of node representation learning, proximity-based embedding models have frequently been build upon or share many parallels with NLP architectures such as the skip-gram model \cite{word2vec} leveraged by deepwalk \cite{deepwalk} and node2vec \cite{node2vec}.
Whilst the representations learnt by such models have proven to be well-suited for a variety of downstream tasks \cite{www_ne_interp}, the extent to which these representations may be at the same time, \textit{interpretable} and not just ``black boxes" has increasingly become a cause for concern \cite{piaggesi2023dine,gogoglou2019interpretability,www_ne_interp,liu2018interpretation}.
To date, research efforts in this direction have largely, \tb{much like in NLP}, focused on latent dimension interpretation which although fruitful in its own right, neglects the interpretation of \textit{inter-node distances} per se, the focus of this work. With a growing number of consumption diversity (see the $GS$ score \cite{gs1,gs2}) and recommendation \cite{word2vec-rec,word2vec-rec2,word2vec-rec3,item2vec} studies now making use of inter-node distances out-of-the-box, being able to interpret their meaning in a simple, comprehensible manor is a more relevant task than ever.
\tb{The present work explores one simple strategy for interpreting inter-node distances: interpretation through a network's $n$-hop proximities. \tb{The \textit{goodness} of a simple interpretation is partially dependent on the existence of a sufficient one-to-one mapping between nodes' proximities in a graph $G$ and closenesses in an embedding space $e$. If such a condition is fulfilled, the relation between between both variables is monotonic and thus, given a closeness score between two nodes it is possible to \textit{interpret} an associated proximity in $G$ without further decoding (and thus, it is simple) given knowledge of the underlying function which relates both variables. Likewise, vice versa from proximity to closeness. 
Building upon this understanding, our work introduces and applies a novel framework to evaluate the interpretability of $e$ by assessing the degree to which there exists a distinct mapping between 1, 2 and higher-order proximities in $G$ and closeness in $e$.} 
Applying four classical node embedding models, deepwalk \cite{deepwalk}, node2vec \cite{node2vec}, SDNE \cite{Wang} and Truncated Singular Value Decomposition (SVD) to two novel music streaming-based artist co-occurrence networks, our work provides a novel account of node embedding models' interpretability which we hope will further inform researchers' model selection processes particularly in domains where interpretation is paramount.}

\section{Related Work}
\subsection{$n$-proximity Semantics}
To interpret {inter-node} distances in vector space we turn our attention to the field of structural linguistics which has inspired the development of many NLP and thus, node proximity-based embedding models. Of particular relevance to this work is the concept of \textit{syntagmatic} and \textit{paradigmatic} relations \cite{parad-dsm} which in more graph theoretic terms, closely corresponds to first- and second-order proximities (\hbox{i.e.} hops) \cite{goyal2017graph}. In linguistics, syntagmatic relations refer to a co-occurrence between two words in a sentence (e.g. `eat',  `food') whilst paradigmatic relations refer broadly to substitution. More precisely, two words are said to be strongly pragmatically connected if they frequently occur in similar contexts, for instance the synonyms ``big'' and ``large''. Inspired by the `distributional semantic hypothesis' \cite{parad-dsm}, classical NLP models such as skip-gram \cite{word2vec} typically try to operationalise paradigmatic relations resulting in their reflection in embedded geometry \cite{word2vec-second-order}. Building upon this understanding, recent works in the fields of explainable manifold learning \cite{HAN2022877} as well as graph embedding evaluation \cite{jin2021understanding} have sought to interpret embedded distances by their respective correspondences to paradigmatic weights by applying measures such as cosine similarity.

Besides having a strong grounding in structural linguistics, paradigmatic weights also closely relate to the notion of \textit{structural similarity} first introduced in the field of social network analysis \cite{lorrain1971structural}, a relaxed notion of the stricter concept of `structural equivalence' applied to segment nodes into sets of equivalent `roles' \cite{Rossi_2015,jin2021understanding}. Closely connected to this work, Jin et al. \cite{jin2021understanding} study correspondences between the similarity of nodes' roles (of which one is structural) and inter-node nearest neighbour distances in embedded space. Our work differs in that we focus exclusively on proximity strengths rather than the more general notion of roles and such, we appraise not just second-order paradigmatic proximites (denoted by Jin et al. as structural roles) but also higher-order proximities \cite{scholkemper}. 
\subsection{Node Embedding Interpretability}
Research concerning the \textit{post-hoc} interpretation of node representations can broadly be conceptualised under two taxonomies: (1) interpretation of single dimensions and (2) interpretation of collections of dimensions (\hbox{i.e.} structural representations). 
Regarding the former, Gogoglou et al. \cite{gogoglou2019interpretability} study how individual dimensions of embedded space tend to correspond to both a network's internal structure and node labels assigned externally. Similarly, Piaggesi et al. \cite{piaggesi2023dine} develop and apply novel measures that capture the global interpretability of embedding vectors based on the marginal contribution of embedding dimensions in predicting a network's structure.
Regarding the latter, Dalmia et al. \cite{www_ne_interp} explore how several node embedding models encode elementary node properties such as Page Rank, node degree, closeness centrality and clustering coefficients through a prediction task. In a similar direction, Liu et al. \cite{liu2018interpretation} leverage the entire structural representation of node embedding models by formulating interpretation as a problem of taxonomy induction from a node's vector representations. More closely connected to this work, Park et al. \cite{park2022providing} introduce a novel framework to interpret node representations through other target nodes in the treated network. Whilst similar in many respects, our research differs in that we focus on interpreting inter-node distances rather than dimensions.

\section{Datasets}
\begin{table}[t]
    \centering
    \small
    \setlength{\tabcolsep}{5pt}
        \begin{tabular}{ccccccccccc}
         
          & $n$-order & $|V|$ & $|E|$ & $Q$ & $Q_{\gamma}$ & $\left\langle W_1 \right\rangle$ & $\left\langle W_2 \right\rangle$ & $\left\langle W_3 \right\rangle$ & $\left\langle W_4 \right\rangle$\\
         \toprule
         \multirow{3}{*}{DeezerPL} & $S$  & 5950 & 306,984 & 0.40 & 0.41 & 1.43 & 3.60 & 5.90 & 8.86\\
         & $P$ & / & 6,564,746 & / & 0.21 & 0.01 & 0.07 & 0.16 & 0.36\\
         & $H$ & / & 10,642,266 & / & 0.18 & 0.01 & 0.13 & 0.41 & 0.84\\
         \midrule
         \multirow{3}{*}{DeezerSess} & $S$ & 2681 & 101,989 & 0.24 & 0.22 & 1.32 & 3.30 & 5.33 & 8.00\\
         & $P$ & / & 2,168,791 & / & 0.09 & 0.01 & 0.05 & 0.13 & 0.35\\
         & $H$ & / & 1,423,672 & / & 0.08 & 0.04 & 0.16 & 0.45 & 0.80\\
         \bottomrule
        \end{tabular}
        \caption{Network properties after imposing the  various filtering criteria. $Q$ and $Q_\gamma$ denote genre modularity scores before and after low-degree filtering respectively (DeezerSess: $\gamma=14$, DeezerPL: $\gamma=15$).}
    \label{tab:dataset_table}
\end{table}
We experiment with embedding two undirected positive point wise mutual information (PPMI) networks of artist co-occurrences constructed from a random sample of anonymised user consumption and curation logs supplied in confidence by the Deezer music streaming platform. {We focus on co-occurrence networks since typically, these are the form of networks embedded in recommendation and consumption diversity studies where inter-node distance measurements are widely applied. \tb{For instance, in the context of nearest neighbour-based recommendation, suggestions of items similar to a given target item are generated by querying its neighbourhood in latent vector space}.\newline\newline
\textbf{DeezerPL: } 
Building upon outstanding efforts treating human-generated playlists as a natural language \cite{Papreja2019,mcfee2011natural,gs2} we focus on a cohort of approximately $1$ million anonymised private Deezer playlists randomly sampled from the platform at the end of 2022. Each playlist is represented by an ordered list of artist ids ($393,624$ unique at the time of extraction), containing an average number of songs of $\avg{|\mathcal{p}|} = 58.32$, $\sigma_\mathcal{p}=139.32$. Upon further examination of this dataset we observe a marked number of extremely long outlier playlists which have a length of up to approximately $10K$ songs. Despite their length, artist redundancy \cite{villermet2021follow} is amongst the highest for these playlists. We theorise that these may be generated by parties other than users and as such, filter playlists where $|\mathcal{p}| > 2\sigma_\mathcal{p}$ to remove these outliers from our analysis set. We then focus on the subset of playlists that have at least $10$ unique artists and thus, contain rich co-occurrence information. After applying both filterings $245,556$ playlists and $285,028$ artists remain. 
Due to computational constraints we then employ a random sampling of $5K$ playlists from our filtered dataset which yields $11,637$ unique artists across all playlists. 
Finally, since our analysis is not concerned with self-directed co-occurrences we reduce each playlist to only contain one occurrence of each of the artists featured in the playlist and construct an artist-artist co-occurrence matrix populated with within-session co-occurrence counts. As is typical in an NLP pipeline we remove unrepeated co-occurrences as these are deemed to be hapaxes and cast our matrix to a Positive Point Wise Mutual Information (PPMI) matrix to account for connectivity due to sheer popularity effects. More formally:
\begin{equation}
S_{i,j} = \max \left(\log _2 \frac{p(i, j)}{p(i) p(j)}, 0\right)
\end{equation}
{where $p(i,j)$ is the proportion of co-occurrences which contain both nodes $i$ and $j$, $p(i)$ denotes the proportion of co-occurrences that involve artist node $i$ and thus, $p(i) p(j)$ denotes the probability of co-occurrence if the within-session occurrence of $i$ and $j$ is assumed to be independent.} This forms our final first-order so-called syntagmatic network $S$. \newline\newline
\textbf{DeezerSess: }
Alongside our playlist-based network we also construct a session-based artist co-occurrence network from $7859$ randomly sampled anonymised Deezer user session logs in the month of January 2018. New sessions of continuous platform engagement are defined when a platform drop off longer than 20 minutes occurs and as per \cite{gs2}, streams lasting less than 30 seconds are filtered as `skips'. Before any further filtering our dataset consists of 396,100 sessions and 29,012 unique artists.  
We proceed by filtering artists who's name (utilised for sense checking the semantics of each embedded space) could not be assigned and then subsequently, filter so-called \textit{fandom} sessions which only contain one artist. Phrased differently, applying both of these filtering strategies means that we only analyse sessions that contain $\ge 2$ artists with assigned names. Removing users who no longer have any sessions after applying this filtering reduces our user set size to $708$. We then impose a further filtering constraint that users must have $>50$ sessions (equating to a lower bound of on-average $1.61$ sessions per day) to focus on users who actively use the platform as a source of music consumption. This reduces our user set to $532$ users who collectively consume $9225$ unique artists across $32,526$ sessions. 
For each user we then compute an artist-artist co-occurrence matrix of size $9225 \times 9225$ which we normalise by the number of sessions they engage in (i.e. the upper bound for a co-occurrence) to account for user activity. We then compute the element-wise summation over all user matrices and cast the summed matrix to a PPMI matrix forming our final network, $S$.\newline\newline
\textbf{Low-Degree Node Filtering: }
As a last processing step we perform a network-level filtering of low-degree nodes which have a degree $< \gamma$ where $\gamma \in \mathbb{N}$ thus, elevating the mechanical effect of low-degree nodes which lack signal to compute node attraction scores (see section \ref{node_attr}). For each network we assign $\gamma$ to be the maximum value possible whilst still ensuring $\leq 50\%$ of nodes are filtered. As depicted in table~\ref{tab:dataset_table}, this filtering strategy does not appear to affect the overall genre-homogeneous community structure typical of music co-occurrence networks \cite{afchar2023spiky} reflected by the minimal variance in genre modularity scores before and after filtering.
After performing this filtering strategy the DeezerPL network consists of 4 connected components, the largest of which spans 5880 nodes and has an average shortest path of 2.35. By contrast the DeezerSess network only consists of one connected component and has an average shortest path of 2.69. 

\section{Node Embedding Models}
We test four classical node embedding models: (1) SVD, (2) DeepWalk, (3) node2vec, (4) SDNE, using the GEM python package \cite{goyal2017graph,goyal3gem}.
Each model (aside from SVD which does not require iterative learning) is trained to output dense 128 dimensional real vectors for 100 epochs.\newline\newline
\textbf{SVD --} 
{As a paradigmatic baseline we apply SVD to $S$, an industry standard employed by music streaming platform to construct user and item embeddings \cite{afchar2023spiky}.}
More formally, SVD is a factorisation technique which seeks to approximate a matrix $A \in \mathbb{R}^{b \times c}$ as the product of three matrices, $B \in \mathbb{R}^{b \times m}$, $\Sigma \in \operatorname{diag}\left(\mathbb{R}_{+}^m\right)$ and $C \in \mathbb{R}^{c \times m}$. 
In practice, the resulting factorisation corresponds to finding an optimal solution which minimises the $\|A-\hat{A}\|_F^2$ loss where $\hat{A}=B \Sigma C^*$, $B$ and $C$ are orthonormal by columns and $\hat{A}$ is a low rank approximation of $A$. In our analysis we make use of the $B \Sigma$ component as our latent item embedding.\newline\newline
\textbf{DeepWalk (dw) \cite{deepwalk} --} The seminal node embedding model leverages the skip-gram \cite{word2vec} architecture from NLP by treating vertex paths traversed by random walks as sentences and the vertices therein as terms to be embedded. 
We focus on tuning 2 hyperparameters -- the walk length ($l$) and context size ($c$) -- by fixing one of the hyperparameters to be assigned to its default value ($l=50, c=10$) and adjusting the other parameter within the following discrete parameter space: $l \in \{20,50,140\}$ and $c \in \{2,6,10,14\}$. For both the DeezerPL and DeezerSess networks we select a final hyperparameter configuration ($c=10$, $l=20$) which yields the minimum {average rank whereby models are ranked by $I$ score (see Sec.~\ref{interp-sec}) for syntagmatic, paradigmatic and higher proximities.\newline\newline
\textbf{node2vec (n2v) \cite{node2vec} --} Similar to dw, n2v performs random walks on a network which later act as input for the skip-gram model. It however differs in its introduction of two additional parameters $p$ and $q$ which bias the random walking process. The former constrains two-hop redundancy whilst the latter biases random walks to be broad or deep. Assigning $p,q=1$ to be both parameters default values we perform a hyperparameter tuning in the range $p,q \in \{1/4, 1, 4\}$ in a manor which is equivalent to that described for dw. 
\tb{Applying the same hyperparameter selection criteria detailed for n2v}  we find that for the DeezerSess network $(p=0.25, q=1)$ yields the best performance whereas a hyperparameter tuning more biased towards a breadth-first exploration $(p=1, q=4)$ is more optimal for the DeezerPL network.\newline\newline 
\textbf{Structural Deep Network Embedding (SDNE) \cite{Wang} --} 
In addition to skip-gram models we also apply SDNE due to its high levels of flexibility regarding syntagmatic and paradigmatic preservation controlled by the $\alpha$ hyperparameter. 
The model is comprised of both an unsupervised deep autoencoder which constructs an embedding from which a vertex neighbourhood can be reconstructed and a supervised part which applies a loss when first-order syntagmatic vertices' latent representations are dissimilar. To explore the effect of introducing syntagmatic-paradigmatic bias variations we construct two versions of the SDNE model: SDNE$_P$ ($\alpha=0.01$) and SDNE$_S$ ($\alpha=0.5$).
\section{Proximity and Attraction}
\subsection{$n$-proximities} 
We first centre our attention on the extent to which a node's neighbourhood typically remains geometrically close or distant in embedding space across different proximities – what we denote as \textit{attraction} (or \textit{repulsion}) dynamics due to the dynamic movement of nodes during embedding training. To this end, we focus on three $n$-order (or  $n$-hops) inter-node proximities, in short \textit{$n$-proximities}: first-order syntagmatic ($n=1$), second-order paradigmatic ($n=2$) and higher-order (with $n=4$) proximities. To evaluate the weight of such proximities we construct three weighted un-directed proximity networks 
which are represented by their adjacency matrices $S, P, H \in \mathbb{R}^{|V|}\times\mathbb{R}^{|V|}$ based on a node set $V$ (see table \ref{tab:dataset_table}).
As previously described, edge weights in our first-order proximity network $S$ simply correspond to PPMI scores between nodes.
To capture second-order paradigmatic proximities we form a structural paradigmatic weighted adjacency matrix $\hat{P}$ by querying node-node distances using cosine similarity such that:
\begin{equation}
\hat{P}_{ij}=
\begin{cases}
\frac{S_i\cdot S_j}{\|S_i\|\|S_j\|}&\text{if}\: i\neq j\\
0&\text{otherwise}
\end{cases}
\end{equation}
whereby self-loops are removed.
We similarly build a higher-order proximity network $\hat{H}$ by computing paradigmatic relations of paradigmatic relations, whereby 2 nodes are connected when they exist in a 4-hop relation in $S$. More formally:
\begin{equation}
\hat{H}_{ij}=
\begin{cases}
\frac{P_i\cdot P_j}{\|P_i\|\|P_j\|}&\text{if}\: i\neq j\\
0&\text{otherwise}
\end{cases}
\end{equation}

We then compute the final versions $P$ and $H$ of the paradigmatic and higher order adjacency matrices by imposing the additional constraint that a relation can only exist in the first order it appears and not beyond (i.e. in absentia \cite{Sahlgren2006}). More formally, $P_{ij}=\hat{P}_{ij}$ if $S_{ij}=0$, and 0 otherwise; likewise, $H_{ij}=\hat{H}_{ij}$ if $P_{ij}=0$, and 0 otherwise. 
For each of our 3 proximity networks we then segment non-zero edges (i.e. node tuples) by their associated edge weights into four sets of increasing proximity -- $(W_1, W_2, W_3, W_4)$ -- by applying k-means with $k=4$. Average weights in each proximity class are depicted in table \ref{tab:dataset_table}. In addition, we compute a further control class $W_0$ which contains all node tuples which do not feature as edges in $S$, $P$ and $H$ (i.e., $S_{i,j} = P_{i,j} = H_{i,j} = 0$). 
\subsection{Neighbourhood Attraction Score}
\label{node_attr}
Given a node embedding model to be evaluated, we first appraise the extent to which typically, a node's $n$-proximity neighbourhoods (i.e. neighbourhoods in $S$, $P$, and $H$) are attracted towards or repulsed from that node in embedded space.
To this end, we define a series of cumulative observation window sizes common to the whole dataset. Denoting $D$ as the set of all pairwise inter-node cosine distances ($1-\cos$) derived from our embedding space, 
we define a series of $|V|+1$ growing window sizes $(w_i)_{i\in\{0,..,|V|\}}$: \begin{equation}w_i = min(D) + \frac{i(\operatorname{max}(D)-\operatorname{min}(D))}{|V|}\end{equation}
thus configuring a uniform window expansion rhythm featuring strides of constant size $\frac{\operatorname{max}(D)-\operatorname{min}(D)}{|V|}$. 
We consider each node in $V$ as a \textit{target node} ($t$) from which we shall compute attractions towards, also examining 
the extent to which node tuples in each weighted proximity set $(W_0, ..., W_4)$ typically exhibit distinct \textit{attraction} dynamics. For each $t\in V$ we denote $N_j= \{v | (t,v)\in W_j\}$ as its $W_j$-filtered neighbourhood. Together $N_j$ are non-overlapping sets spanning over $V \setminus \{t\}$.
The pipeline that follows is detailed for the $N_1$ neighbourhood however equivalent logic applies for all other neighbourhoods $N_j$. We first assert that $N_1$ contains a sufficient number of nodes for plotting and curve fitting that will follow by imposing the filter that $|N_1| > 0.005\cdot|V|$.
Given that this condition holds we then compute a cumulative \textit{hit} vector of size $|V|$ whereby each element corresponds to the proportion of the neighbourhood captured in window $w_i$:
\tb{\begin{equation}
    h_i = \frac{|\{v\in N_1 : d(t,v) < w_i\}|}{|N_1|}
\end{equation}}
In other words, the $h$ vector describes the growing amount of nodes of the $N_1$ neighbourhood which remain within a growing radius around $t$.
For curve fitting purposes we then associate this $h$ vector with an $x$ vector whereby $x_i = -6 + \frac{12(i-1)}{|V|}$ and fit a sigmoid function on $x$ to $h$ using non-linear least squares:
\begin{equation}
f(x) = \frac{1}{1+e^{-g(x-s)}}
\end{equation}
whereby $g$ and $s$ are the parameters to be fitted that respectively affect growth and shift of the sigmoid. Using $f$ we then compute node $t$'s neighbourhood attraction score \hbox{w.r.t.} nodes in $N_1$ simply as:
\begin{equation}
    \delta_1 = \int_{-6}^6 f(x)
\end{equation}
%
\begin{figure}[t]
    \centering
    \includegraphics[width=0.9\textwidth]{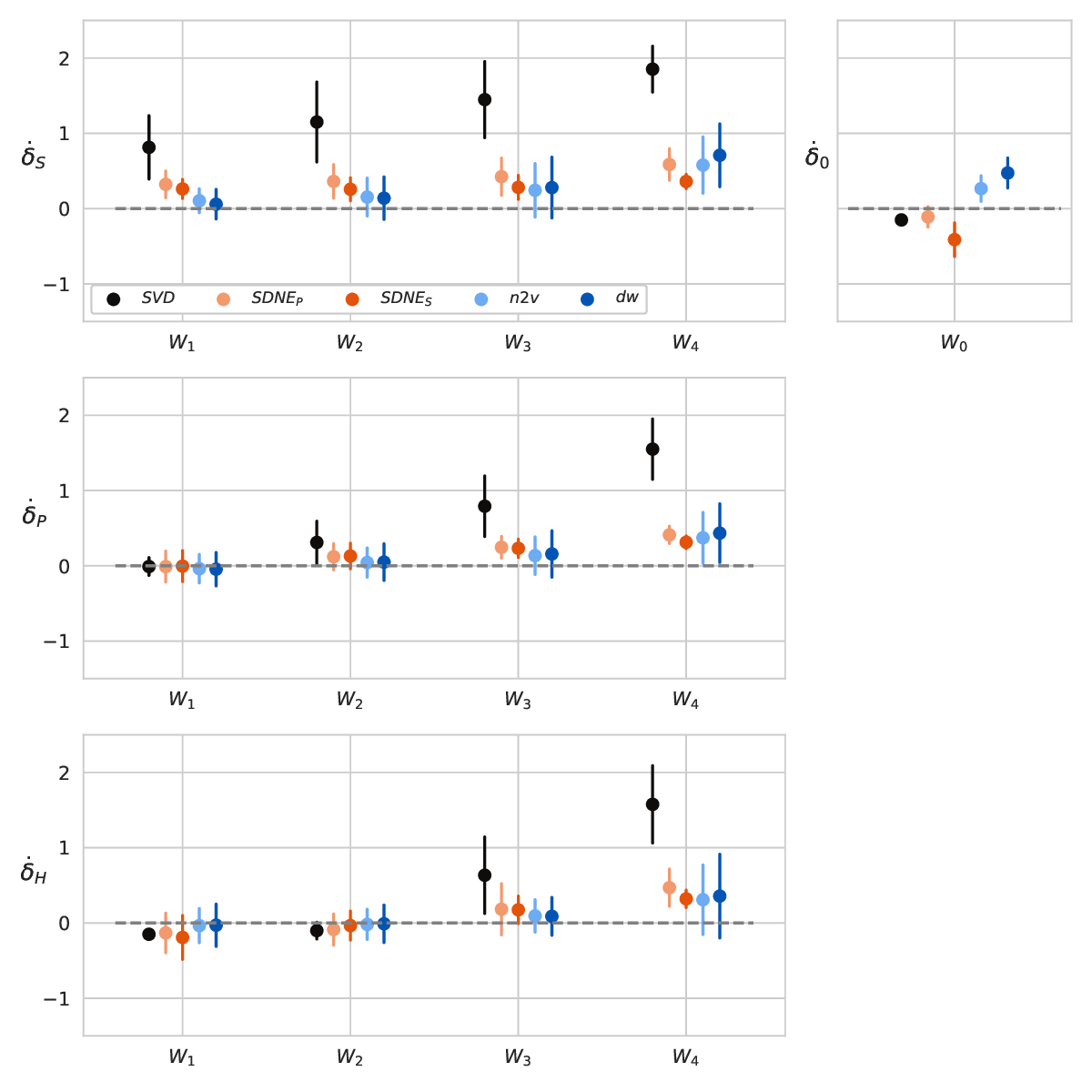}
    \caption{Attraction scores for the DeezerSess network before z-normalisation for first-order ($\dot{\delta}_S$), second-order ($\dot{\delta}_P$), higher-order ($\dot{\delta}_H$) and other ($\dot{\delta}_0$) proximities which acts as a control set.}
    \label{fig:delta_fig_sess}
\end{figure}
Greater values indicate stronger local attraction dynamics to the target node $t$ (i.e., a bigger area under the cumulative hit curve), which denotes a better preservation of closeness in embedded space. Inversely, smaller values reflect stronger local repulsion dynamics away from $t$ (i.e., a smaller area under the curve).
To elicit comparisons of $\delta_1$ between different models which may display unique item density distributions in vector space we normalise $\delta_1$ by the neighbourhood attraction score  ($\tilde{\delta}$) of a null model which reflects the average $h$ over all nodes irrespective of the proximity class:
\begin{equation*}
    \dot{\delta_1} = \operatorname{log}_2(\delta_1 / \tilde{\delta})
\end{equation*}
and likewise for all other $N_j$ and thus, $\delta_j$ and $\dot{\delta_j}$.
\subsection{Results}
We first centre our attention on attraction differences between neighbourhoods which exist in distinct $n$-proximities (i.e. between $S$, $P$ and $H$). As depicted in Figures \ref{fig:delta_fig_sess} and \ref{fig:delta_fig_pl}, lower-order proximities (i.e., $S$) tend to experience on-average stronger attraction dynamics. This effect is particularly pronounced for SVD reflected in the high $\delta$ scores obtained even for the weakest neighbourhood ($N_1$).
\begin{figure}[t]
    \centering
    \includegraphics[width=0.9\textwidth]{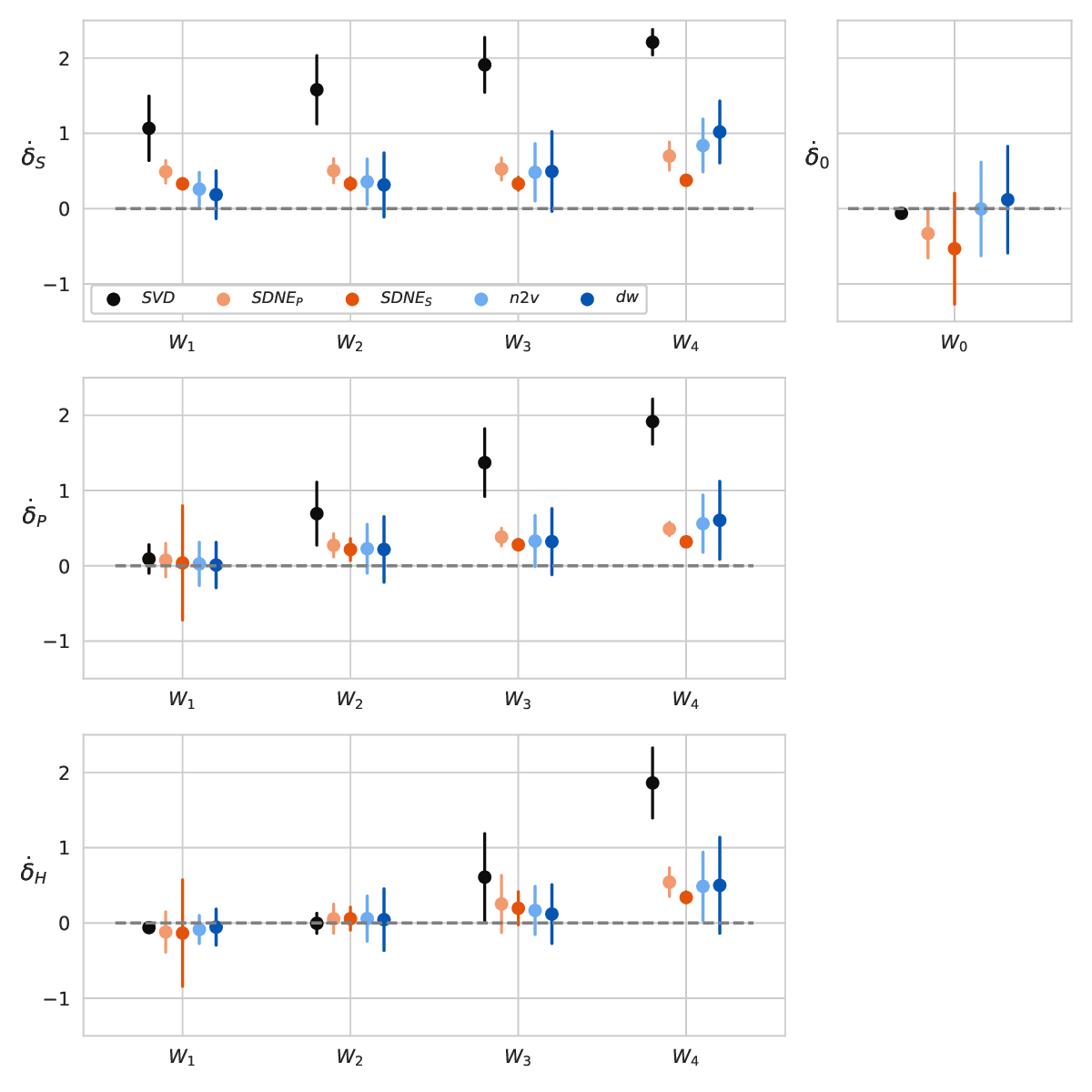}
    \caption{Attraction scores for the DeezerPL network before z-normalisation for first-order ($\dot{\delta}_S$), second-order ($\dot{\delta}_P$), higher-order ($\dot{\delta}_H$) and other ($\dot{\delta}_0$) proximities which acts as a control set.}
    \label{fig:delta_fig_pl}
\end{figure}
Regarding model differences, all models aside from dw and n2v display greater attractions to 1-, 2- and 4-proximities than other proximities reflected by the control set $W_0$. SVD consistently exhibits the highest attraction scores across all $n$-proximities tested even in the case of neighbourhoods in higher proximity relations despite implicitly optimising structural 2-order relations. Indeed, it appears that SVD is capable of capturing both low and high-order relations in its geometry and, furthermore, distinguishing between different $H$ neighbourhoods as we shall later demonstrate. 
Regarding the model's capacity to consistently locally position 1-order proximities we hypothesise that this effect may be due, rather, to the typical structure of co-occurrence networks which exhibit a strong correlation between $S$ and $P$ proximities (average $S$, $P$ correlation over all nodes is $r=0.643$ for DeezerSess and $r=0.628$ for DeezerPL). Hence, optimising $P$ relations can implicitly result in $S$ optimisation by proxy. 
Nonetheless, the SDNE$_S$ model which, by design, tries to closely embed nodes in strong $S$ relations tends in fact, to display weaker first-order attraction dynamics than its second-order paradigmatic counterpart (SDNE$_P$). Interestingly it appears that, within the context of co-occurrence network embedding, trying explicitly to operationalise first-order relations may be both redundant and furthermore, counter-productive.
\section{Appraising $n$-proximity Interpretability}
\label{interp-sec}
The interpretability of an embedding model in a simple manor depends in part on the existence of a sufficient one-to-one mapping between proximity weights (from an $n$-order adjacency matrix) and embedding distances. For instance, a strong (network) proximity should not correspond to the same (embedding) distance as a weak proximity. To quantify this, we propose to compute the dissimilarity between attraction score distributions associated with each proximity class ($W_j$), the main focus of this work. In order to interpret a given inter-node distance with confidence, proximity class mappings should each be associated with embedding distance mappings that are as distinct as possible from one another.
For all $t \in V$ we segment valid $\dot{\delta}$ into 5 sets of attraction scores -- $\Delta_0$, $\Delta_1$, $\Delta_2$, $\Delta_3$, $\Delta_4$ which correspond to attraction scores for $N_0$, $N_1$, $N_2$, $N_3$, $N_4$ neighbourhoods respectively.
To compare between models whose attraction scores span different ranges we perform a form of z-normalisation that allows us to place each model's attraction scores on a comparable scale whilst still capturing attraction variations between each $W_j$ as well as absolute differences between trends inter-nodes.
\begin{equation}
    \tilde{\Delta}_{i} = \{(\dot{\delta} - \left\langle \mu_{\Delta} \right\rangle) / \sigma_{\mu_{\Delta}} : \dot{\delta} \in \Delta_i \}
\end{equation}
where $\mu_{\Delta} = (\left\langle \Delta_0 \right\rangle, \left\langle \Delta_1 \right\rangle, \left\langle \Delta_2 \right\rangle, \left\langle \Delta_3 \right\rangle, \left\langle \Delta_4 \right\rangle)$.

Binning each $\tilde{\Delta}_i$ into a discrete histogram representation with a bin size of $\sigma_{\mu_{\Delta}}/ 4$ in the range $[-10\sigma_{\mu_{\Delta}},..., 10\sigma_{\mu_{\Delta}}]$, we construct the following count normalised histograms $P = (P_{0}, P_{1}, P_{2}, P_{3}, P_{4})$. We then compute our final interpretability score, $I$ as:
\begin{equation}
I = \frac{\sum_{i=1}^{|P|-1} \sum_{j=i+1}^{|P|} JS(P_i, P_j)}{\binom{|P|}{2}}
\end{equation}
where $JS$ reflects the symmetric Jensen-Shannon score noting $m$ the pointwise mean of $p$ and $q$, and $KL$ the Kullback-Leibler divergence, $$JS(p,q)=\sqrt{\frac{1}{2}(KL(p \| m)+KL(q \| m)})$$
To summarise, we compute z-normalised attraction score sets $\tilde{\Delta}_j$ for each neighbourhood $N_j$, which we bin into 80 one-standard-deviation regularly-spaced bins of size one-quarter-of standard deviation. This yields five histograms which we compare across pairs of neighbourhoods $N_j$. High values of $I$ correspond to more distinct pairs $P_i$, $P_j$ and thus, under this conceptualisation a higher interpretability.
\paragraph{Results.}
\newcommand{\sg}[1]{{\tiny ({\tiny$\pm$}#1)}}
\begin{table}[!b]
  \centering 
  \scriptsize
\setlength{\tabcolsep}{3.5pt}
\begin{tabular}{cccccccccccc}
      & \multicolumn{5}{c}{\footnotesize DeezerSess} && \multicolumn{5}{c}{\footnotesize DeezerPL}\\
      &  $\:\:$SVD$\:\:$            & SDNE$_P$  & SDNE$_S$  & $\:\:$n2v$\:\:$   & $\quad$dw$\quad$ &          
      &  $\:\:$SVD$\:\:$            & SDNE$_P$  & SDNE$_S$  & $\:\:$n2v$\:\:$   & $\quad$dw$\quad$ \\
\cmidrule{2-6}\cmidrule{8-12}
$I_S$ & \textbf{*0.69} & *0.57 & *0.53 & *0.57 & *0.60 &
      & \textbf{*0.75} & *0.56 & 0.46 & *0.54 & *0.51 \\
      & \sg{0.16} &  \sg{0.25} & \sg{0.28} & \sg{0.17} & \sg{0.18} &
      & \sg{0.12} &  \sg{0.30} & \sg{0.33} & \sg{0.17} & \sg{0.19}\\

$I_P$ & \textbf{*0.80} & *0.69 & *0.72 & *0.60 & *0.58 &
      & \textbf{*0.77} & *0.74 & *0.70 & *0.51 & *0.46 \\
      & \sg{0.05} &  \sg{0.13} & \sg{0.13} & \sg{0.16} & \sg{0.18} &
      & \sg{0.09} &  \sg{0.11} & \sg{0.17} & \sg{0.13} & \sg{0.12} \\

$I_H$ & \textbf{*0.73} & *0.57 & *0.71 & *0.55 & *0.54 &
      &  0.60 & *0.67 & \textbf{*0.70}& *0.53 & *0.43 \\
      & \sg{0.19} &  \sg{0.22} & \sg{0.13} & \sg{0.20} & \sg{0.23} &
      & \sg{0.33} &  \sg{0.15} & \sg{0.13} & \sg{0.13} & \sg{0.13}\\
\cmidrule{2-6}\cmidrule{8-12}
\end{tabular}
  \caption{Interpretability results for both networks across $S$, $P$ and $H$ proximities (standard deviations in parentheses). Bold values mark maximums per proximity and * indicate that all pairwise comparisons between neighbourhood histograms for a model were significant using a two-sample Kolmogorov-Smirnov test ($\alpha=0.05$ and Bonferroni correction applied).}
  \label{table}
\end{table}
Regarding interpretability differences between $n$-proximities we observe that as depicted in table \ref{table}, aside from dw and n2v who display their highest $I$ scores largely within 1-order proximities, the node embedding models we test have the highest interpretability scores within $2$-order $P$ proximities. Even weaker neighbourhood classes (i.e. $N_1$, $N_2$) are found to display distinct attraction dynamics with respect to 2-order proximities. 
Regarding model differences, just as we saw previously that the dw and n2v models were outliers with respect to neighbourhood attraction dynamics likewise, are they distinct with regard to interpretability. Aside from $S$ proximities, both models are found to yield the lowest $I_P$ and $I_H$ scores.
By contrast, SVD is the most interpretable model across almost all $n$-proximities with a slight exception for the DeezerPL dataset whereby the SDNE$_S$ model surprisingly, attains the highest $I_H$ score hinting at a non-trivial, complex relationship between the optimisation of first- and higher-order proximities. 
Overall, our findings suggest that despite being less state-of-the-art, inter-node distances are consistently the most interpretable via $n$-proximities for the SVD model. We suggests that thus, this model may be well-fitted for application domains where distance interpretation is both paramount and requires minimal computational efforts for instance, when generating on-line interpretable nearest neighbour recommendations from a given seed item. At the same time, we remark that SDNE$_S$ remains a fruitful alternative if both the interpretability and attraction of $S$ proximities are rather counter-intuitively, not a priority.
\section{Discussion}
The contribution of this work introduces and applies a novel pipeline to study node embedding interpretability at the rarely exploited level of inter-node distances. Drawing influence from structural linguistics which inspired the development of several NLP and thus, node embedding models our work appraises the extent to which (embedded) closeness measurements derived from several classical node embedding models' representation spaces tend to correspond to distinct (network) proximities.
In turn, our findings lead us to a conclusion often cited in industry \cite{afchar2023spiky} but yet surprisingly understated in academia: SVD is both the most interpretable model and furthermore, capable of capturing higher order relations within its geometry. In applications such as generating personalised recommendation or user diversity assessments where inter-node distances are applied out-of-the box and thus, required to be reflective of some semantic meaning in their own right, we suggest that the usage of SVD would be advisable. 
At the same time, our work also provides a valuable insight for researchers engaging in co-occurrence network embedding, a common practice in recommender system studies: optimising first-order proximities may be both redundant due to implicit correlations to second-order proximities and furthermore, sub-optimal to increase first-order attractions as was observed to be the case for the SDNE$_S$ model. 
%
%
Whilst beyond the scope of this work, we suggest a fruitful direction for future works may be to expand upon the cohort of models and networks tested in this study to add further rigour to these findings. In addition, another direction may be found in exploring the effect of other distance measures (in the case of this work cosine distance) upon $n$-proximity interpretation and furthermore, the interplay this has with the embedding model deployed.


%
%
%
\bibliographystyle{splncs04}
\bibliography{bib_complenet}

\begin{thebibliography}{10}
\providecommand{\url}[1]{\texttt{#1}}
\providecommand{\urlprefix}{URL }
\providecommand{\doi}[1]{https://doi.org/#1}

\bibitem{afchar2023spiky}
Afchar, D., Hennequin, R., Guigue, V.: Of spiky {SVDs} and music recommendation. In: Proc. 17th RecSys ACM Conf. on Recommender Systems. pp. 926--932 (2023)

\bibitem{gs2}
Anderson, A., Maystre, L., Anderson, I., Mehrotra, R., Lalmas, M.: Algorithmic effects on the diversity of consumption on spotify. In: Proceedings of The Web Conference 2020. p. 2155–2165. WWW '20, Association for Computing Machinery, New York, NY, USA (2020)

\bibitem{item2vec}
Barkan, O., Koenigstein, N.: Item2vec: Neural item embedding for collaborative filtering. arXiv:1603.04259  (2016)

\bibitem{word2vec-rec}
Caselles-Dupr{\'e}, H., Lesaint, F., Royo-Letelier, J.: Word2vec applied to recommendation: Hyperparameters matter. In: Proc. 12th ACM Conf on Recommender Systems RecSys'18. pp. 352--356 (2018)

\bibitem{www_ne_interp}
Dalmia, A., Gupta, M.: Towards interpretation of node embeddings. In: Companion Proceedings of the The Web Conference 2018. pp. 945--952 (2018)

\bibitem{gogoglou2019interpretability}
Gogoglou, A., Bruss, C.B., Hines, K.E.: On the interpretability and evaluation of graph representation learning. arXiv:1910.03081  (2019)

\bibitem{goyal3gem}
Goyal, P., Ferrara, E.: Gem: A python package for graph embedding methods. Journal of Open Source Software  \textbf{3}(29), ~876 (2018)

\bibitem{goyal2017graph}
Goyal, P., Ferrara, E.: Graph embedding techniques, applications, and performance: A survey. Knowledge-Based Systems  \textbf{151},  78--94 (2018)

\bibitem{node2vec}
Grover, A., Leskovec, J.: Node2vec: Scalable feature learning for networks. In: Proc. 22nd ACM SIGKDD Intl Conf on Knowledge Discovery and Data Mining. p. 855–864. Association for Computing Machinery, New York, NY, USA (2016)

\bibitem{HAN2022877}
Han, H., Li, W., Wang, J., Qin, G., Qin, X.: Enhance explainability of manifold learning. Neurocomputing  \textbf{500},  877--895 (2022)

\bibitem{jin2021understanding}
Jin, J., Heimann, M., Jin, D., Koutra, D.: Towards understanding and evaluating structural node embeddings (2021)

\bibitem{parad-dsm}
Lapesa, G., Evert, S., Schulte~im Walde, S.: Contrasting syntagmatic and paradigmatic relations: Insights from distributional semantic models. In: Proc. Third Joint Conf. on Lexical and Computational Semantics (*{SEM} 2014). pp. 160--170. Association for Computational Linguistics and Dublin City University, Dublin, Ireland (Aug 2014)

\bibitem{liu2018interpretation}
Liu, N., Huang, X., Li, J., Hu, X.: On interpretation of network embedding via taxonomy induction. In: Proc. 24th ACM SIGKDD Intl Conf on Knowledge Discovery \& Data Mining. pp. 1812--1820 (2018)

\bibitem{lorrain1971structural}
Lorrain, F., White, H.C.: Structural equivalence of individuals in social networks. The Journal of mathematical sociology  \textbf{1}(1),  49--80 (1971)

\bibitem{mcfee2011natural}
McFee, B., Lanckriet, G.R.: The natural language of playlists. In: ISMIR. vol.~11, pp. 537--541 (2011)

\bibitem{word2vec}
Mikolov, T., Chen, K., Corrado, G., Dean, J.: Efficient estimation of word representations in vector space. In: Bengio, Y., LeCun, Y. (eds.) 1st Intl Conf on Learning Representations, {ICLR} 2013, Scottsdale, Ariz., USA, May 2-4 (2013)

\bibitem{word2vec-rec2}
Musto, C., Semeraro, G., Degemmis, M., Lops, P.: Word embedding techniques for content-based recommender systems: An empirical evaluation. In: RecSys Posters. vol.~1441 (2015)

\bibitem{word2vec-rec3}
Ozsoy, M.G.: From word embeddings to item recommendation. arXiv:1601.01356  (2016)

\bibitem{Papreja2019}
Papreja, P., Venkateswara, H., Panchanathan, S.: Representation, exploration and recommendation of playlists. In: Cellier, P., Driessens, K. (eds.) Machine Learning and Knowledge Discovery in Databases - Intl Workshops of ECML PKDD 2019. pp. 543--550. Springer, Switzerland (2020)

\bibitem{park2022providing}
Park, H.: Providing post-hoc explanation for node representation learning models through inductive conformal predictions. IEEE Access  \textbf{11},  1202--1212 (2022)

\bibitem{deepwalk}
Perozzi, B., Al-Rfou, R., Skiena, S.: Deepwalk: Online learning of social representations. In: Proc 20th ACM SIGKDD Intl Conf on Knowledge discovery and data mining. pp. 701--710 (2014)

\bibitem{piaggesi2023dine}
Piaggesi, S., Khosla, M., Panisson, A., Anand, A.: Dine: Dimensional interpretability of node embeddings. arXiv:2310.01162  (2023)

\bibitem{Rossi_2015}
Rossi, R.A., Ahmed, N.K.: Role discovery in networks. {IEEE} Transactions on Knowledge and Data Engineering  \textbf{27}(4),  1112--1131 (apr 2015)

\bibitem{Sahlgren2006}
Sahlgren, M.: The Word-Space Model: Using distributional analysis to represent syntagmatic and paradigmatic relations between words in high-dimensional vector spaces. Ph.D. thesis, Institutionen f{\"o}r lingvistik (2006)

\bibitem{word2vec-second-order}
Schlechtweg, D., Oguz, C., Schulte~im Walde, S.: Second-order co-occurrence sensitivity of skip-gram with negative sampling. In: Proc. 2019 ACL Workshop BlackboxNLP: Analyzing and Interpreting Neural Networks for NLP. pp. 24--30. Association for Computational Linguistics, Florence, Italy (Aug 2019)

\bibitem{scholkemper}
Scholkemper, M., Schaub, M.T.: Local, global and scale-dependent node roles. In: 2021 IEEE International Conference on Autonomous Systems (ICAS). pp.~1--5 (2021)

\bibitem{villermet2021follow}
Villermet, Q., Poiroux, J., Moussallam, M., Louail, T., Roth, C.: Follow the guides: disentangling human and algorithmic curation in online music consumption. In: Proc. 15th ACM Conf. on Recommender Systems. pp. 380--389 (2021)

\bibitem{gs1}
Waller, I., Anderson, A.: Generalists and specialists: Using community embeddings to quantify activity diversity in online platforms. In: Proceedings of The Web Conference 2019. p. 1954–1964. WWW '19, Association for Computing Machinery, New York, NY, USA (2019)

\bibitem{Wang}
Wang, D., Cui, P., Zhu, W.: Structural deep network embedding. In: Proc. 22nd ACM SIGKDD Intl Conf on Knowledge Discovery and Data Mining. p. 1225–1234. Association for Computing Machinery, New York (2016)

\end{thebibliography}

\end{document}